# Specific Heat Discontinuity, ΔC, at $T_c$ in $BaFe_2(As_{0.7}P_{0.3})_2$ – Consistent with Unconventional Superconductivity

J. S. Kim[1], G. R. Stewart[1], S. Kasahara[2], T. Shibauchi[3], T. Terashima[2], and Y. Matsuda[3]

[1]Department of Physics, University of Florida
Gainesville, FL 32611-8440
[2] Research Center for Low Temperature and Materials Sciences,
Kyoto University, Sakyo-ku, Kyoto 606-8501, Japan
[3] Department of Physics, Kyoto University,
Sakyo-ku, Kyoto 606-8502, Japan

**Abstract:** We report the specific heat discontinuity, $\Delta C/T_c$, at $T_c = 28.2$ K of a collage of single crystals of $BaFe_2(As_{0.7}P_{0.3})_2$ and compare the measured value of 38.5 mJ/molK$^2$ with other iron pnictide and iron chalcogenide (FePn/Ch) superconductors. This value agrees well with the trend established by Bud'ko, Ni and Canfield who found that $\Delta C/T_c \propto aT_c^2$ for 14 examples of doped $Ba_{1-x}K_xFe_2As_2$ and $BaFe_{2-x}TM_xAs_2$, where the transition metal TM=Co and Ni. We extend their analysis to include all the FePn/Ch superconductors for which $\Delta C/T_c$ is currently known and find $\Delta C/T_c \propto aT_c^{1.9}$ and a=0.083 mJ/molK$^4$. A comparison with the elemental superconductors with $T_c$>1 K and with A-15 superconductors shows that, contrary to the FePn/Ch superconductors, electron-phonon-coupled conventional superconductors exhibit a significantly different dependence of ΔC on $T_c$, namely $\Delta C/T_c \propto T_c^{0.9}$. However $\Delta C/\gamma T_c$ appears to be comparable in all three classes (FePn/Ch, elemental and A-15) of superconductors with, e. g., $\Delta C/\gamma T_c$=2.4 for $BaFe_2(As_{0.7}P_{0.3})_2$. A discussion of the possible implications of these phenomenological comparisons for the unconventional superconductivity believed to exist in the FePn/Ch is given.

## Introduction

In the ongoing study of superconductivity in the FePn/Ch, only a few examples have been found where the preponderance of the evidence argues for nodes in the superconducting gap. Among these, overdoped $BaFe_{2-x}Co_xAs_2$, $T_c$ =8.1 K, is believed to be a nodal superconductor based on thermal conductivity as a function of magnetic field, $\kappa$(H), data.[1] Based on penetration depth [2], nuclear magnetic resonance [3], $\kappa$(H) [2], and as yet unpublished specific heat vs H data[4], $BaFe_2(As_{0.7}P_{0.3})_2$ is also believed to be a nodal superconductor.

Herein we report measurements of the discontinuity in the specific heat, $\Delta C$, divided by $T_c$ in $BaFe_2(As_{0.7}P_{0.3})_2$ at the superconducting transition temperature ($T_c^{mid}$=28.2 K) which, due to the high temperature and necessity for precision measurement because of the large phonon contribution (>90% of $C_{Total}$), has not previously been reported. Bud'ko, Ni and Canfield (hereafter 'BNC') have pointed out [5], in a sampling of 14 doped $BaFe_2As_2$ FePn superconductors, that $\Delta C/T_c$ has an unusual dependence on $T_c$, namely $\propto T_c^2$. Our data for $BaFe_2(As_{0.7}P_{0.3})_2$ are discussed in this framework; as well, an expansion of this discussion to include conventional electron-phonon coupled and several representative heavy Fermion superconductors is included for comparison. Finally, since the specific heat $\gamma$ (limit of C/T as $T\rightarrow 0$) of $BaFe_2(As_{0.7}P_{0.3})_2$ is known from measurements [4] of the specific heat up to 35 T (2/3 of $H_{c2}(0)$), the normalized $\Delta C/\gamma T_c$ is also discussed and compared with other FePn/Ch and conventional superconductors where $\gamma$ has been reported.

## Experimental

The specific heat of a 40 mg piece of high purity (99.9985 %) platinum from Alfa Aesar, which has approximately the same specific heat as the 18.2 mg collage of crystals of $BaFe_2(As_{0.7}P_{0.3})_2$ at 30 K, was first measured using established time constant methods.[6] In the temperature range of measurement, the measured specific heat of the Pt standard was within ±1% of published values. The single crystals used for the collage of $BaFe_2(As_{0.7}P_{0.3})_2$ were prepared as discussed in [7], and were the same as used in our other specific heat measurements.[4,8]

**Results and Discussion**

The specific heat of $BaFe_2(As_{0.7}P_{0.3})_2$ from 19 to 32 K is shown in Fig. 1. The transition

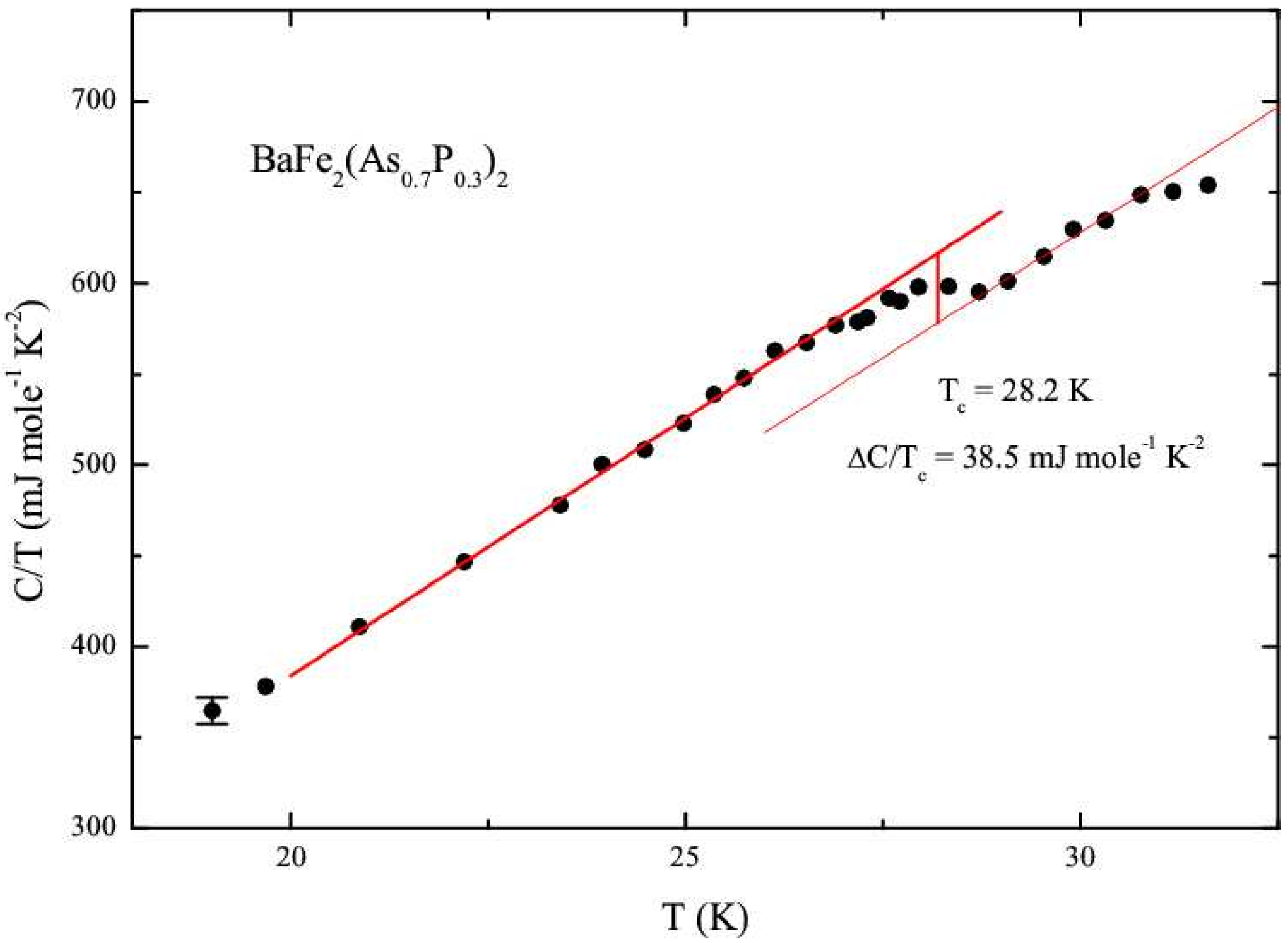

Fig. 1 (color online) Specific heat in the neighborhood of the superconducting transition in $BaFe_2(As_{0.7}P_{0.3})_2$. The precision and the absolute error of the data, ±1%, is approximately the size of the data points. The sloped red lines are fits to the data above and below the transition, while the vertical red line marks an ideal, sharp discontinuity at $T_c^{mid}$=28.2 K.

width of ≈1.3 K for the bulk specific heat measurement is not dissimilar to the magnetic susceptibility result[3] although the $T_c^{onset}$ determined from susceptibility is approximately 30 K.

The measured $\Delta C/T_c$ is 38.5 ± 2 mJ/molK$^2$, which agrees well with the trend in $\Delta C/T_c$ vs $T_c$ established by BNC and is compared to other FePh/Ch superconductors in Fig. 2. Bud'ko, Ni and Canfield [5] presented data for their specific heat results on Co- and Ni-doped $BaFe_2As_2$,

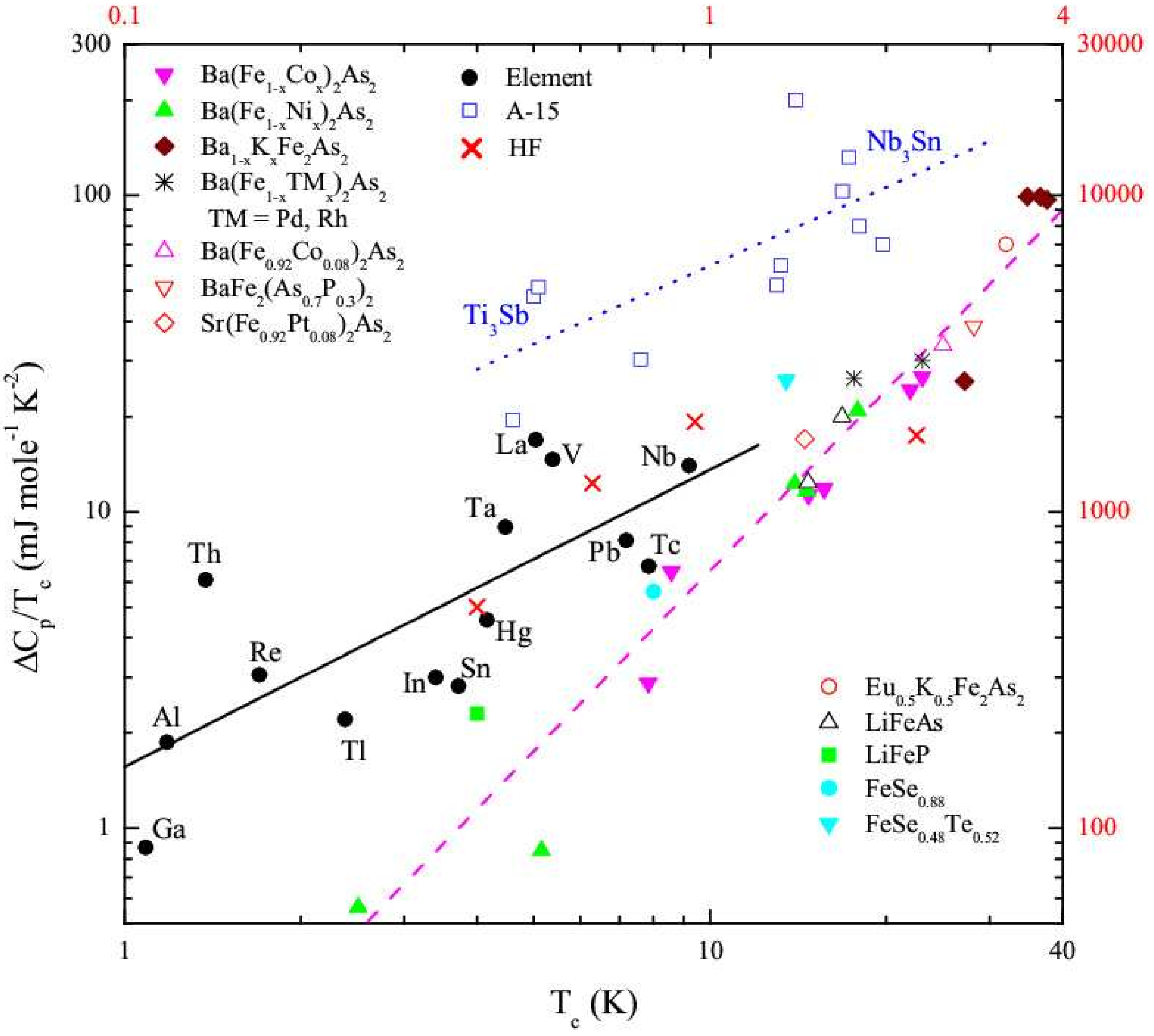

Fig. 2 (color online) $\log\Delta C/T_c$ vs $\log T_c$ for conventional superconductors (elements with $T_c$>1 K in black solid circles and representative A-15 compounds in blue open squares), for FePn/Ch superconductors, and for some representative heavy Fermion superconductors (red X's) which, due to the lower $T_c$'s and higher $\Delta C$'s involved are plotted vs the red right hand and upper axes with different scales. The slope of the black elemental superconductor line gives $\Delta C/T_c \propto T_c^{0.94}$, for the A-15 superconductors the dotted blue best fit line gives $\Delta C/T_c \propto T_c^{0.82}$, while the magenta dashed line for the FePn/Ch gives $\Delta C/T_c \propto 0.083T_c^{1.89}$. A-15 $Nb_3Ge$, which has a rather broad transition that is difficult to extrapolate to an idealized, sharp transition, is not plotted. In order to focus on the FePn/Ch results, the labels for the A-15 and heavy Fermions are mostly omitted. However, Table 1 lists the detailed data values and references for all the plotted points. The heavy Fermion compounds, which are likely unconventional superconductors, surprisingly show $\Delta C/T_c$ vs $T_c$ behavior similar to the elemental and A-15 superconductors, with $\Delta C/T_c$ roughly varying as $T_c^{0.7}$. This type of plot, with data for the early results in FePn for doped $BaFe_2As_2$, was first proposed by Bud'ko, Ni, and Canfield.[5]

as well as literature data for $Ba(Fe_{0.939}Co_{0.061})_2As_2$, $Ba_{0.55}K_{0.45}Fe_2As_2$ and $Ba_{0.6}K_{0.4}Fe_2As_2$, see Fig. 2. In addition, we show more recent data on other FePn/Ch, all the superconducting elements with $T_c$>1 K, a broad selection of A-15 superconductors in order to extend the $T_c$ range higher to overlap more with the FePn/Ch, and several heavy Fermion superconductors to also provide a comparison to known unconventional superconductors. The respective $\Delta C/T_c$ and $T_c$ values are shown in Table 1. The gamma values for the elemental superconductors are bounded by around 10 mJ/molK$^2$ (V and La) [6], while $\gamma$ values for the A-15's can be several times larger (see references in Table 1). The slopes of the two $\Delta C/T_c$ vs $T_c$ sets of data in Fig. 2 for these two classes of conventional, electron-phonon coupled superconductors are clearly quite close, and in strong contrast to the slope for the FePn/Ch.

The data shown in Fig. 2 for the FePn/Ch are complete with three exceptions. The value [9-10] (not shown) for $KFe_2As_2$, where for a high quality (residual resistivity ratio ≈ 650) sample [10] $\Delta C/T_c \approx 41$ mJ/molK$^2$ at $T_c^{mid}$=3.1 K, lies far above the FePn/Ch behavior and seems consistent with the A-15 line. This deviation - together with other properties of $KFe_2As_2$ such as lack of Fermi surface nesting, atypical pnictogen height vs $T_c$ behavior, and low $H_{c2}$ (for a review, see [11]) - argues that superconductivity in $KFe_2As_2$ may be different from that observed in the other FePn/Ch shown in Fig. 2. For LaFePO, the result [12] that $\Delta C/T_c \approx 3$ mJ/molK$^2$, $T_c^{mid} \approx 5.5$ K (which agrees with the FePn/Ch trend shown in Fig. 2) is also not shown in Fig. 2 since not only is the sample only partially superconducting (C/T does not trend to 0 below $T_c$), but at least one work [13] has argued that stoichiometric LaFePO is non-superconducting. Recently, the first $\Delta C/T_c$ data have become available [66] for the new $K_xFe_{2-y}Se_2$, $T_c \approx 32$ K, superconductor which – in contrast to the other FePn/Ch superconductors – appears to lack hole pockets at the Fermi surface as well as being much less metallic. Whether the large deviation in this new material's $\Delta C/T_c$, reported to be $\approx 12$ mJ/molK$^2$, from the BNC trend shown in Fig. 2 is intrinsic or due to sample quality must await further work and is therefore not included in the plot here.

Zaanen [14] has proposed that this $\Delta C/T_c \propto T_c^2$ scaling behavior argues against a Fermi liquid picture, and instead discusses the idea that the superconductivity could be forming from a non-Fermi liquid quantum critical metal. Rather than the usual [15] quantum critical *point* in a phase diagram, Zaanen argues for a quantum critical *region* over some fraction of the superconducting dome in composition space. It is certainly true that several FePn/Ch show non-Fermi liquid behavior in their resistivities, e. g. P-doped $BaFe_2As_2$ [7] (whose $\Delta C/T_c$ is shown in Fig. 1), Co-doped $BaFe_2As_2$ [16], and FeSe [17] exhibit $\rho=\rho_0+AT^1$ while LiFeAs shows [18] $\rho=\rho_0+ AT^{1.5}$. However, the question of non-Fermi liquid behavior is open for the great majority of the FePn/Ch whose $\Delta C/T_c$'s are summarized in Fig. 2. To explain the observed BNC scaling Kogan [19] considers instead that the FePn/Ch superconductors are weak coupled Fermi liquids with strong pair breaking, with the observed $\Delta C$'s and $T_c$'s much reduced from those in hypothetical clean material.

Looking at the trends brought out in Fig. 2 in a more phenomenological way, this sort of BNC plot makes clear that whatever the pairing mechanism in the superconducting state in the FePn/Ch is, that this superconductivity is different in a fundamental fashion from conventional superconductivity. Conventional electron-phonon coupled elemental and A-15 superconductors have a $\Delta C/T_c$ that is dependent [20] on three factors: the electronic density of states at the Fermi energy, N(0), the spectral density $\alpha^2F(\omega)$ and the Coulomb pseudopotential $\mu^*$. This dependence, using the slopes of the fits of $\Delta C/T_c$ to $T_c^{\alpha}$ in Fig. 2, says that for such superconductors a.) these three factors combine to give $\Delta C/T_c \propto T_c$ and b.) since in conventional superconductors experimental values for $\Delta C/T_c$ roughly vary [21] as $\gamma$ ($\Delta C/\gamma T_c$=1.43 in weak coupling Bardeen Cooper Schrieffer, BCS, theory), $T_c$ then (broadly speaking) must vary as $\gamma$, which is proportional to $N(0)(1+\lambda_{el\text{-}ph})$. Thus, for conventional elemental and A-15 superconductors we find in Fig. 2 that $\Delta C/T_c \propto T_c$ and - since experimentally [21] and according to BCS theory $\Delta C/T_c \propto \gamma$ - therefore $T_c \propto \gamma \propto N(0)(1+\lambda_{el\text{-}ph})$. (In weak coupling BCS theory, $T_c \propto \exp(-1/N(0)V)$.) This roughly linear phenomenological relation between $T_c$ and the renormalized density of states in conventional superconductors that results from the plot in Fig. 2 is of course the paradigm that drove the search for higher $T_c$ in the A-15 superconductors, with some success. It is also the paradigm that Bednorz and Mueller ignored to discover unconventional high $T_c$ superconductivity in the low density of state perovskite-structure cuprates.

Now, the BNC plot, updated in Fig. 2 in the present work, suggests another paradigm, namely that whatever instead of (or in addition to) N(0), $\alpha^2F(\omega)$, and $\mu^*$ determines $\Delta C/T_c$ for the FePn/Ch, the result is that $\Delta C/T_c$ varies approximately as $T_c^2$ in these new superconductors.

Another difference for the FePn/Ch is that although $\Delta C/T_c$ – in so far as $\gamma$ values are known – remains [22] approximately proportional to $\gamma$, the measured $\gamma$'s combined with calculations imply that $\gamma$ for the FePn/Ch comes primarily from $N(0)(1+\lambda_{el-el})$ since $\lambda_{el-ph}$ is negligible [23]. Thus, since for the FePn/Ch $\Delta C/T_c \propto T_c^2$ and $\Delta C/T_c \propto \gamma \propto N(0)(1+\lambda_{el-el})$, the BNC plot has implications for how the superconducting transition temperature $T_c$ in FePn/Ch depends on the *electron-electron interactions* that are presumably involved in the superconducting pairing.

It is also interesting to note the behavior, see Fig. 2 and Table 1, for several heavy Fermion superconductors: $CeIrIn_5$ and $CeCoIn_5$ (both non-Fermi liquid systems and both believed to have unconventional superconductivity (d-wave gap for $CeCoIn_5$), see Pfleiderer [61]), $CeCu_2Si_2$ and $UBe_{13}$. Their $\Delta C/T_c$ vs $T_c$ slope shows a similar behavior to the conventional elemental and A-15 superconductors, thus the FePn/Ch present *another kind* of unconventional superconductivity than the heavy Fermion's. The further question – what about $\Delta C/T_c$ vs $T_c$ for the cuprates – runs into complications caused by the pseudogap behavior with, e. g., various compositions near x=0.2 in $La_{1-x}Sr_xCuO_4$ having [65] similar $T_c$'s and much different $\Delta C$'s.

**Summary and Conclusions:**

The specific heat jump at the superconducting transition, $\Delta C$, has been measured for $BaFe_2(As_{0.7}P_{0.3})_2$. Normalized as $\Delta C/T_c$, the result (38.5 ± 2 mJ/molK$^2$) follows the phenomenological result of Bud'ko, Ni and Canfield (BNC) for 14 doped $BaFe_2As_s$ that $\Delta C/T_c$ varies approximately as $T_c^2$. The present work has widened the comparison to include eight other FePn/Ch and finds that the BNC $\Delta C/T_c \propto aT_c^{\alpha}$ ($\alpha$=1.89±0.1, a=0.083 mJ/molK$^{3.89}$) correlation still holds. In order to investigate this correlation, comparisons to elemental and A-15 superconductors, as well as the unconventional heavy Fermion superconductors, were presented. Both the conventional superconductors and the heavy Fermion superconductors exhibit an entirely different dependence, namely $\Delta C/T_c \propto T_c^{0.7\text{-}0.9}$, i. e. the FePn/Ch are clearly following a different dependence than these other superconductors. A heuristic argument, based on the facts that a.) *all* of the aforementioned superconductors appear to approximately follow $\Delta C/\gamma T_c \propto$ constant and b.) the specific heat $\gamma$ ($\equiv C/T$ as $T \rightarrow 0$) is proportional to the electronic density of states at the Fermi energy, N(0), times $(1+\lambda_{el\text{-}ph}+\lambda_{el\text{-}el})$, is advanced that the different dependence of $\Delta C/T_c$ on $T_c$ in the FePn/Ch yields information on the dependence of $T_c$ on $\lambda_{el\text{-}el}$ in these iron containing superconductors.

Acknowledgements: Work at Florida performed under the auspices of the US Department of Energy, contract no. DE-FG02-86ER45268. Helpful conversations with Professor Ilya Vekhter are gratefully acknowledged.

**Table 1 $\Delta C/T_c$, $T_c$ and references for elemental and A-15 superconductors, FePn/Ch superconductors, and heavy Fermion superconductors**

| Superconductor element/A-15 | $\Delta C/T_c$ | $T_c$ | Ref. | Superconductor FePn/Ch, heavy Fer | $\Delta C/T_c$ | $T_c$ | Ref. |
|---|---|---|---|---|---|---|---|
| | (mJ/molK$^2$) | (K) | | | (mJ/molK$^2$) | (K) | |
| Ga | 0.87 | 1.087 | 24 | $Ba(Fe_{1-x}Co_x)_2As_2$, x=0.038 | 3 | 7 | 5 |
| Al | 1.87 | 1.18 | 25 | x=0.047 | 12 | 15 | 5 |
| Th | 6.1 | 1.374 | 26 | x=0.058 | 27 | 22.6 | 5 |
| Re | 3.06 | 1.70 | 27 | x=0.061 | ≈35 | 24 | 47 |
| Tl | 2.21 | 2.38 | 28 | x=0.078 | 24 | 22 | 9 |
| In | 3 | 3.4 | 29 | x=0.10 | 12 | 15.5 | 5 |
| Sn | 2.82 | 3.718 | 29 | x=0.114 | 6.5 | 8.5 | 5 |
| Hg | 4.56 | 4.16 | 28 | $Ba(Fe_{1-x}Ni_x)_2As_2$, x=0.024 | 0.5 | 2.5 | 5 |
| Ta | 8.95 | 4.47 | 30 | x=0.032 | 12 | 15 | 5 |
| La | 16.9 | 5.04 | 31 | x=0.046 | 21 | 18 | 5 |
| V | 14.63 | 5.379 | 32 | x=0.054 | 12.5 | 14 | 5 |
| Pb | 8.14 | 7.19 | 33 | x=0.072 | 1 | 5 | 5 |
| Tc | 6.75 | 7.86 | 34 | $Ba_{0.55}K_{0.45}Fe_2As_2$ | 23 | 30 | 48 |
| Nb | 14 | 9.2 | 35 | $Ba_{0.6}K_{0.4}Fe_2As_2$ | 100 | 34.6 | 49 |
| $Mo_{3.2}Pt_{0.8}$ | 19.5 | 4.6 | 36 | $Ba_{0.6}K_{0.4}Fe_2As_2$ | 98 | 36.5 | 50 |
| $Ti_3Sb$ | 48 | 5 | 37 | $Ba_{0.6}K_{0.4}Fe_2As_2$ | 100 | 37.3 | 51 |
| $Ti_3Ir_{0.8}Pt_{0.2}$ | 52 | 5 | 38 | $Ba(Fe_{0.943}Rh_{0.057})_2As_2$ | 31 | 22.8 | 52 |
| $V_3Re$ | 32 | 7.8 | 39 | $Ba(Fe_{0.957}Pd_{0.043})_2As_2$ | 23 | 17.5 | 53 |
| $Nb_3Au_{0.7}Pt_{0.3}$ | 52 | 13 | 40 | $Ba(Fe_{0.92}Co_{0.08})_2As_2$ | 33.6 | 25 | 53 |
| $Mo_{1.6}Tc_{2.4}$ | 60 | 13.2 | 41 | $Sr(Fe_{0.92}Pt_{0.08})As_2$ | 17 | 14.5 | 54 |
| $V_3Ga$ | 200 | 14 | 42 | $Eu_{0.5}K_{0.5}Fe_2As_2$ | 70 | 32 | 55 |
| $V_3Si$ | 103 | 16.8 | 43 | LiFeAs | 12.4 | 14.7 | 56 |
| $Nb_3Sn$ | 122 | 17.8 | 44 | LiFeAs | 20 | 16.8 | 57 |
| $Nb_3Al$ | 80 | 18 | 45 | LiFeP | 2.3 | 4 | 58 |
| $Nb_3Al_{0.8}Ge_{0.2}$ | 70 | 19.7 | 46 | $FeSe_{0.88}$ | 5.6 | 8 | 59 |
| | | | | $FeSe_{0.48}Te_{0.52}$ | 23 | 13.5 | 60 |
| | | | | $CeIrIn_5$ | 500 | 0.4 | 61 |
| | | | | $CeCoIn_5$ | 1740 | 2.25 | 62 |
| | | | | $CeCu_2Si_2$ | 1230 | 0.63 | 63 |
| | | | | $UBe_{13}$ | 1920 | 0.94 | 64 |

## References

[1] Dong J K, Zhou S Y, Guan T Y, Qiu X, Zhang C, Cheng P, Fang L, Wen H H and Li S Y 2010 *Phys. Rev.* B **81** 094520

[2] Hashimoto K, Yamashita M, Kasahara S, Senshu Y, Nakata N, Tonegawa S, Ikada K, Serafin A, Carrington A, Terashima T, Ikeda H, Shibauchi T and Matsuda Y 2010 *Phys. Rev.* B **81** 220501(R)

[3] Nakai Y, Iye T, Kitagawa S, Ishida K, Kasahara S, Shibauchi T, Matsuda Y and Terashima T 2010 *Phys. Rev.* B **81** 020503(R)

[4] Kim J S et al. unpublished.

[5] Bud'ko S L, Ni N and Canfield, P C 2009 *Phys. Rev.* B **79** 220516(R)

[6] Stewart G R 1983 *Rev. Sci. Instrum.* **54** 1

[7] Kasahara S, Shibauchi T, Hashimoto K, Ikada K, Tonegawa S, Okazaki R, Ikeda H, Takeya H, Hirata K, Terashima T and Matsuda Y 2010 *Phys. Rev.* B **81** 184519

[8] Kim J S, Hirschfeld P J, Stewart G R, Kasahara S, Shibauchi T, Terashima T and Matsuda Y 2010 *Phys. Rev.* B **81** 214507

[9] Fukazawa H, Yamada Y, Kondo K, Saito T, Kohori Y, Kuga K, Matsumoto Y, Nakatsuji S, Kito H, Shirage P M, Kihou K, Takeshita N, Lee C.-H., Iyo A and Eisaki H 2009 *J. Phys. Soc. Japan* **78** 083712

[10] Kim J S, Kim E G, Stewart G R, Chen X H and Wang X F, Phys. Rev. B **83**, 172502 (2011).

[11] Johnston D C 2010 *Adv. in Physics* **59** 803

[12] Analytis J G, Chu J-H, Erickson A S, Kucharczyk C, Serafin A, Carrington A, Cox C, Kauzlarich S M, Hope H and Fisher I R arXiv0810.5368; Suzuki S, Miyasaka S, Tajima S, Kida T and Hagiwara, M *J. Phys. Soc. Japan* **78** 114712

[13] McQueen T M, Regulacio M, Williams A J, Huang Q, Lynn J W, Hor Y S, West D V, Green M A and Cava R J 2008 *Phys. Rev.* B **78** 024521

[14] Zaanen J 2009 *Phys. Rev.* B **80** 212502

[15] Stewart G R 2001 *Rev. Mod. Phys.* **73** 797

[16] Ahilan K, Balasubramaniam J, Ning F L, Imai T, Sefat A S, Jin R, McGuire M A,

Sales B C and Mandrus D 2008 *J. Phys.: Condens. Matter* **20** 472201

[17] Sidorov V A, Tsvyashchenko A V and Sadykov R A arXiv0903.2873; Masaki S, Kotegawa H, Hara Y, Tou H, Murata K, Mizuguchi Y and Takano Y 2009 *J. Phys. Soc. Japan* **78** 063704

[18] Lee B S, Khim S H, Kim J S, Stewart G R and Kim K H 2010 *Europhysics Letters* **91** 67002; Khim S H and Kim K H, 2010, private communication

[19] Kogan V G 2009 *Phys. Rev.* B **80** 214532

[20] Carbotte J P *Rev. Mod. Phys.* **62** 1027

[21] Using the $\Delta C/T_c$ values for the superconducting elements shown in Fig. 2/Table 1 and the $\gamma$ values for the elements [6], $\Delta C/\gamma T_c$ varies between 1.34 for Re to 2.45 for Hg and 2.72 for Pb. For the A-15 superconductors, $\Delta C/\gamma T_c$ varies between 1.5 for $Ti_3Sb$ to about 2.1 for $Nb_3Al$, see the respective references given in Table 1

[22] $\gamma$ values to match with the $\Delta C/T_c$ values for the FePn/Ch are somewhat rare, since for $T_c$'s above 10 K the extrapolation of the normal state C/T to T=0 to obtain $\gamma$ ($\equiv$C/T for T$\rightarrow$0) is difficult. In addition to the value of $\Delta C/\gamma T_c$ for the present work of 38.5/16=2.4 (gamma from [4, 8], values of $\gamma$ are known for $Ba_{0.6}K_{0.4}Fe_2As_2$ ($\gamma$=49 mJ/molK$^2$, Kant Ch, Deisenhofer J, Günther A, Schrettle F, Loidl A, Rotter M and Johrendt D 2010 *Phys. Rev.* B **81** 014529) to give $\Delta C/\gamma T_c$ from 2.04 to 2.36 and for annealed optimally doped $BaFe_{1.85}Co_{0.16}As_2$ ($\gamma$=22 mJ/molK$^2$, Gofryk K, Sefat A S, McGuire M A, Sales B C, Mandrus D, Imai T, Thompson J D, Bauer E D and Ronning F 2011 *J. Phys.: Conf. Ser.* **273** 012094) to give $\approx$1.5 for $\Delta C/\gamma T_c$. Thus, for FePn/Ch systems where $\gamma$ has been determined, $\Delta C/\gamma T_c$ is comparable to that found for conventional superconductors.

[23] Boeri L, Dolgov O V and Golubov A A 2009 *Physica* C **469** 628; Subedi A, Zhang L, Singh D J and Du M H 2008 *Phys. Rev.* B **78** 134514

[24] Seidel G and Keesom P H 1958 *Phys. Rev.* **112** 1083

[25] Hopkins D C 1962 PhD thesis University of Illinois unpublished

[26] Luengo C A, Cotignola J M, Sereni J G, Sweedler A R, Maple M B and Huber J G 1972 Solid State Comm. **10** 459

[27] Smith D R and Keesom P H 1970 *Phys. Rev.* B **1** 188

[28] van der Hoeven B J C and Keesom P H 1964 *Phys. Rev.* **135** A631

[29] Bryant C A and Keesom P H 1961 *Phys. Rev.* **123** 491

[30]  Cochran J F 1962 *Annals of Physics* **19** 186

[31]  Pan P H, Finnemore D K, Bevolo A J, Shanks H R, Beaudry B J, Schmidt F A and Danielson G C 1980 *Phys. Rev.* **21** 2809

[32]  Radebaugh R and Keesom P H 1966 *Phys. Rev.* **149** 209

[33]  Neighbor J E, Cochran J F and Shiffman C A 1967 *Phys. Rev.* **155** 384

[34]  Trainor R J and Brodsky M B 1975 *Phys. Rev.* B **12** 4867

[35]  Leupold H A and Boorse H A 1964 *Phys. Rev.* **134** A1322

[36]  Flükiger R, Paoli A, Roggen R, Yvon K and Muller J *Solid State Comm.* **11** 61

[37]  Junod A, Heiniger F, Muller J and Spitzli P 1970 *Helv. Phys. Acta* **43** 59

[38]  Junod A, Flükiger R and Muller J 1976 *J. Phys. Chem. Solid* **37** 27

[39]  Giorgi A L, Matthias B T and Stewart G R 1978 *Solid State Comm.* **27** 291

[40]  Spitzli P 1971 *Phys. Kondens. Materie* **13** 22

[41]  Stewart G R and Giorgi A L 1980 *J. Low Temp. Phys.* **41** 73

[42]  Morin F J, Maita J P, Williams H J, Sherwood R C, Wernick J H and Kuntzler J E 1962 *Phys. Rev. Lett.* **8** 275

[43]  Stewart G R and Brandt B L 1984 *Phys. Rev.* B **29** 3908

[44]  Stewart G R, Cort B and Webb G W 1981 *Phys. Rev.* B **24** 3841; Wieland L J and Wicklund A W 1968 *Phys. Rev.* **166** 424

[45]  Cort B, Stewart G R, Snead C L, Sweedler A R and Moehlecke S 1981 *Phys. Rev.* B **24** 3794; Willens R H, Geballe T H, Gossard A C, Maita J P, Menth A, Hull G W and Soden R R 1969 *Solid State Comm.* **7** 837

[46]  Stewart G R, Szklarz E G and Giorgi A L 1978 *Sol. State Comm.* **28** 5

[47]  Chu J-H, Analytis J G, Kucharczyk C and Fisher I R 2009 Phys. Rev. **79** 014506

[48] Ni N, Bud'ko S L, Kreyssig A, Nandi S, Rustan G E, Goldman A I, Gupta S, Corbett J D, Kracher A and Canfield P C 2008 *Phys. Rev.* B **78** 014507

[49] Welp U, Xie R, Koshelev A E, Kwok W K, Luo H Q, WangZ S, Mu G and Wen H H 2009 *Phys. Rev.* B **79** 094505

[50] Mu G, Luo H, Wang Z, Shan L, Ren C and Wen H H 2009 *Phys. Rev.* B **79** 174501

[51] Rotter M, Tegel M, Schellenberg I, Schappacher F M, Pöttgen R, Deisenhofer J, Günther A, Schrettle F, Loidl A and Johrendt D 2009 *N. Journal of Physics* **11** 025014

[52] Ni N, Thaler A, Kracher A, Yan J Q, Bud'ko S L and Canfield P C 2009 *Phys. Rev.* B **80** 024511

[53] Gofryk K, Sefat A S, McGuire M A, Sales B C, Mandrus D, Imai T, Thompson J D, Bauer E D and Ronning F 2011 *J. Phys.: Conf. Ser.* **273**, 12094

[54] Kirshenbaum K, Saha S R, Drye T and Paglione J 2010 *Phys. Rev.* B **82** 144518

[55] Jeevan H S and Gegenwart P 2010 *J. Phys.: Conf. Series* **200** 012060

[56] Stockert U, Abdel-Hafiez M, Evtushinsky D V, Zabolotnyy V B, Wolter A U B, Wurmehl S, Morozov I, Klingeler R, Borisenko S V and Büchner B arXiv1011.4246

[57] Lee B S, Khim S H, Kim J S, Stewart G R and Kim K H 2010 *Europhys. Lett.* **91** 67002

[58] Deng Z, Wang X C, Liu Q Q, Zhang S J, Lv Y X, Zhu J L,Yu R C and Jin C Q 2009 *Europhys. Lett.* **87** 37004

[59] Hsu F-C, Luo J-Y, Yeh K-W, Chen T-K, Huang T-W, Wu P M, Lee Y-C, Huang Y-L, Chu Y-Y, Yan D-C and Wu M-K 2008 *Proc. Nat. Acad. Sci.(USA)* **105** 14262

[60] Braithwaite D, Lapertot G, Knaf W and Sheikin I 2010 *J. Phys. Soc. Japan* **79** 053703

[61] Pfleiderer C 2009 *Rev. Mod. Phys.* **81**, 1551

[62] Kim J S, Stewart G R and Bedorf D 2009 *J. Low Temp. Phys.* **157**, 29

[63] Stewart G R 1984 *Rev. Mod. Phys.* **56** 755

[64] Scheidt E-W, Schreiner T, Kumar P and Stewart G R 1998 *Phys. Rev.* B **58** 15153

[65] Loram J W, Luo J, Cooper J R, Liang W Y and Tallon J L 2001 *J. Phys. Chem. Sol.* **62** 59

[66] Zeng B, Shen B, Chen G F, He J B, Wang D M, Li C H and Wen H H 2011 *Phys. Rev.* B **83** 144511